%% file: main.tex
\documentclass[Afour,sagev,times]{sagej}

\usepackage{moreverb,url}
\usepackage[colorlinks,bookmarksopen,bookmarksnumbered,citecolor=red,urlcolor=red]{hyperref}
\newcommand\BibTeX{{\rmfamily B\kern-.05em \textsc{i\kern-.025em b}\kern-.08em
T\kern-.1667em\lower.7ex\hbox{E}\kern-.125emX}}

\def\volumeyear{2022}

\usepackage{xcolor}
\graphicspath{{fig/}}
\usepackage{subcaption}
\usepackage{glossaries}
\glsdisablehyper
\loadglsentries{acronyms}
\usepackage{listings}

\begin{document}

\runninghead{Risco-Martín et al.}

\title{Simulation-driven engineering for the management of harmful algal and cyanobacterial blooms}

\author{José L. Risco-Martín\affilnum{1} and Segundo Esteban\affilnum{1} and Jesús Chacón\affilnum{1} and Gonzalo Carazo-Barbero\affilnum{1} and Eva Besada-Portas\affilnum{1} and José A. López-Orozco\affilnum{1}}

\affiliation{\affilnum{1}Universidad Complutense de Madrid, Spain}

\corrauth{José L. Risco-Martín, Department of Computer Architecture and Automation,
Universidad Complutense de Madrid,
C/Prof. José García Santesmases, 9,
28040 Madrid,
Spain.\\
Tel. +34-91-3947602}

\email{jlrisco@ucm.es}

\begin{abstract} 
\input{0_abstract}
\end{abstract}

\glsresetall


\maketitle

\input{1_introduction}

\input{2_related_work}

\input{3_system_architecture}

\input{4_experiments}

\input{5_conclusion}

\begin{acks}
The authors would like to thank Mr. Giordy Alexander Andrade Aimara, who implemented the integration of actual sensors into DEVS-BLOOM as part of his master's thesis. This work has been supported by the Research Projects IA-GES-BLOOM-CM (Y2020/TCS-6420) of the Synergic program of the Comunidad Autónoma de Madrid, SMART-BLOOMS (TED2021-130123B-I00) funded by MCIN/AEI/10.13039/501100011033 and the European Union NextGenerationEU/PRTR, and INSERTION (PID2021-127648OB-C33) of the  Knowledge Generation Projects program of the Spanish Ministry of Science and Innovation.
\end{acks}

\bibliographystyle{SageV}
\bibliography{biblio}

\begin{biogs}
\textbf{José L. Risco-Martín} received his Ph.D. from UCM, where he currently is Full Professor in the Department of Computer Architecture and Automation. His research interests include systems modeling, simulation, and optimization.

\noindent\textbf{Segundo Esteban} is an Associate Professor of Systems Engineering and Automation at UCM. He holds a Ph.D. in Physics from the same University. His research interests include Systems Modeling and Control.

\noindent\textbf{Jesús Chacón} is an Assistant Professor in the Faculty of Physics Sciences at Universidad Complutense de Madrid. He holds a Ph.D. in Physics from the same University. His research interests include modeling, simulation and implementation of event-based control systems, and education in control engineering.

\noindent\textbf{Gonzalo Carazo-Barbero} is a PhD student in Faculty of Informatics at Universidad Complutense de Madrid. His research interests include modeling and simulation of cyanobacterial blooms and unmanned surface vehicles path planning.

\noindent\textbf{Eva Besada-Portas} is an Associate Professor of Systems Engineering and Automation at UCM. She also holds a PhD in Computer Systems from UCM. Her research interests include uncertainty modeling and simulation, optimal control and planning of unmanned vehicles.

\noindent\textbf{José A. López-Orozco} is a Full Professor in the UCM. He holds a Ph.D. in Physics from the same University. His research interests include multisensor data fusion, control and planning of unmanned vehicles, and robotics.
\end{biogs}


\end{document}

%% file: acronyms.tex
\newacronym{api}{API}{Application Programming Interface}
\newacronym{cps}{CPS}{Cyber-Physical Systems}
\newacronym{devs}{DEVS}{Discrete Event System Specification}
\newacronym{dt}{DT}{Digital Twin}
\newacronym{eems}{EEMS}{EE Modeling System}
\newacronym{ews}{EWS}{Early-Warning System}
\newacronym{gcs}{GCS}{Ground Control Station}
\newacronym{gui}{GUI}{Graphical User Interface}
\newacronym{hab}{HAB}{Harmful Algal and Cyanobacterial Bloom}
\newacronym{iot}{IoT}{Internet of Things}
\newacronym{ms}{M\&S}{Modeling and Simulation}
\newacronym{mbse}{MBSE}{Model Based Systems Engineering}
\newacronym{usv}{USV}{Unmanned Surface Vehicle}

%% file: 0_abstract.tex
\glspl{hab}, occurring in inland and maritime waters, pose threats to natural environments by producing toxins that affect human and animal health. In the past, \glspl{hab} have been assessed mainly by the manual collection and subsequent analysis of water samples and occasionally by automatic instruments that acquire information from fixed locations. These procedures do not provide data with the desirable spatial and temporal resolution to anticipate the formation of \glspl{hab}. Hence, new tools and technologies are needed to efficiently detect, characterize and respond to \glspl{hab} that threaten water quality. It is essential nowadays when the world's water supply is under tremendous pressure because of climate change, overexploitation, and pollution.
This paper introduces DEVS-BLOOM, a novel framework for real-time monitoring and management of \glspl{hab}. Its purpose is to support high-performance hazard detection with \gls{mbse} and \gls{cps} infrastructure for dynamic environments.

%% file: 1_introduction.tex
\section{Introduction}\label{sec:intro}

\glspl{hab} constitute an especially relevant public health hazard and ecological risk, due to their frequent production of toxic secondary metabolites. Exposure to cyanotoxins, for instance, can cause severe health effects in humans and animals, as well as significant economic losses in local communities. 

\glspl{hab} typically emerge in a variety of freshwater ecosystems like reservoirs, lakes, and rivers \cite{Vincent2009}. Their intensity and frequency have increased globally during the last decade, mainly due to the current vulnerability of water resources to environmental changes, such as global warming, population growth, and eutrophication. For example, in 2014, a Microcystis \gls{hab} at the water treatment plant intake for Toledo (Ohio, USA) caused the distribution of non-potable water for more than 400000 people during multiple days \cite{Schmale2019}. The danger is not limited to the closest water environment since extracellular material from freshwater \glspl{hab} has been observed in the water and the atmosphere at locations far beyond their edges.

During the last 30 years, the data needed to estimate the health of a water body and the possible existence of \glspl{hab} have been obtained by specialized personnel through manual collection of water samples and subsequent analysis in the laboratory, and, in the best cases, by automatic instruments placed at fixed locations, that acquire data and, in very few cases, samples. Financial and personnel resource restrictions reduce the manual collection to the moments of the year when \glspl{hab} are more likely to appear at a few geographical points and with minimal frequencies. The delay suffered by analytical results and the limited capacity to interpret the current scenario reduces the reaction (prediction, prevention, and mitigation) capability of the authorities responsible for the distribution of drinking water and its recreational uses \cite{Meriluoto2017}. This is critical when deploying \glspl{ews}, whose essential work is to collect water samples and identify the cyanobacterial cell or algae density as soon as possible. Hence, it is crucial to develop new cost-effective monitoring and early detection systems capable of predicting and anticipating when and where \glspl{hab} form and produce toxins to provide support to water managers/authorities for guiding their policies and protecting the public health through the deployment of effective \glspl{ews}.

In this context, \gls{ms} can be used to clarify the dynamics of \glspl{hab}, as it has historically done in similar areas\cite{Ung1972}. Numerical-based and data-driven machine learning models have been extensively used to simulate \glspl{hab} in aquatic systems \cite{Long2011, Pyo2019}. These techniques try to reach accurate predictions through what we call \textit{base models}. These models have been integrated into more generic software tools like the \gls{eems} \cite{EEMS2022}. Based on these models and tools, various countries have attempted to build \glspl{ews} with the support of predictive systems \cite{Wu2022}. 

\begin{figure*}
  \centering
  \includegraphics[width=0.9\textwidth]{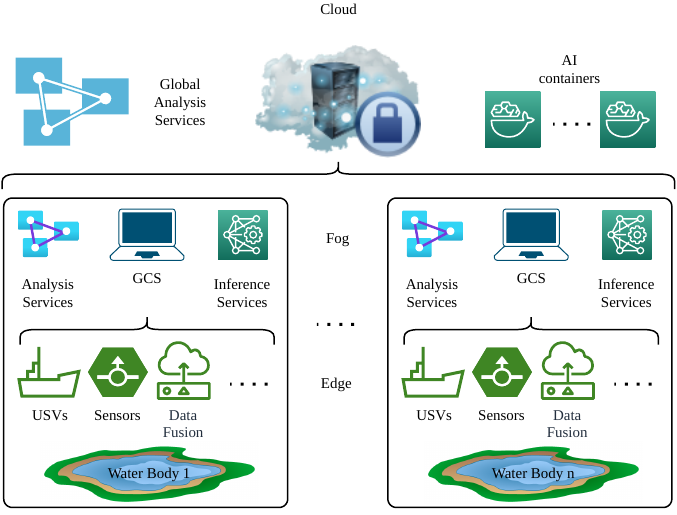}
  \caption{\label{fig:big_picture}Conceptual model of the proposed framework.}
\end{figure*}

Our vision is, however, oriented to a system of systems architecture, a more holistic and \textit{integrative model} that includes not only the use of the aforementioned \textit{base models} but also the infrastructure of the \gls{ews}. Figure \ref{fig:big_picture} shows our conception of the simulation framework, tightly coupled to the represented \gls{cps}. As Figure \ref{fig:big_picture} illustrates, our framework follows a \gls{iot}-based architecture through the use of \glspl{dt}. Water bodies are monitored in the edge layer by a set of sensors, including those onboard automated boats, called here-after \glspl{usv}, that continuously send data to the server at the nearest \gls{gcs} in the fog layer. There, domain experts can analyze the data, run some models, tests, or plan the \glspl{usv} trajectories. The framework supports horizontal scalability, being able to add more water bodies with the support of a cloud layer, where authorities can compare different reports and make high-level decisions.

To simulate and operate this complex model, in this paper we propose DEVS-BLOOM, a novel \gls{ms} framework to enable real-time monitoring and hazard prediction of \glspl{hab}. Our approach is based on the principles of \gls{mbse}: (i) model-based since \gls{mbse} is based on the use of models to represent and manage information about a system, (ii) system-centric, focusing on the system as a whole, (iii) iterative and incremental process, which involves the development of models over time, (iv) collaboration between stakeholders, including system engineers, domain experts, etc., (v) traceability between requirements, design, and implementation, (vi) reuse of models, components, and other artifacts to improve efficiency and reduce the risk of errors, and (vii) verification and validation to ensure that the system meets its requirements and that it operates as intended \cite{Wymore2018}. At the same time, we aim to provide high-performance real-time services, such as detecting outliers or executing complex forecasting methods. All this is achieved through the implementation of model-driven technologies and infrastructure based on the \gls{iot} and \glspl{dt} paradigms. As a result, we address three main topics in the sustainable management of water resources under the umbrella of model-driven technologies: (i) provide a robust interface to design intelligent \glspl{hab} management system prototypes, (ii) provide vertical scalability, modeling the whole pyramidal structure, from the sensors to the authorities, and (iii) provide horizontal scalability, being able of adding more sensors and water bodies with the support of well-grounded \gls{ms} methodologies.

The main contributions of this work can be summarized as follows:

\begin{itemize}
\item We present a framework where we can model the water body, the infrastructure needed to monitor and manage \glspl{hab} like sensors or \glspl{usv}, the computing resources needed to control that infrastructure like workstations or cloud servers, and the actions performed by the human team like operators, domain experts, or water authorities.
\item The model can be simulated in virtual mode to analyze the viability of the whole system; in hybrid mode, where some components are virtual, and others like actual sensors are co-simulated to test or calibrate these sensors; or in real mode, where the framework is not a simulator but a fully operational toolkit, where all the components are real.
\item The framework supports horizontal scalability, allowing us to incorporate more water bodies, or vertical scalability, allowing us to elaborate more complex models. This is possible with the parallel or distributed execution of the framework, which the internal libraries automatically provide.
\end{itemize}

DEVS-BLOOM has been developed through the \gls{devs} \cite{Zeigler2018}, a well known \gls{ms} formalism. To prove the feasibility of each scenario, the framework uses formal models. It can be fed with authentic or synthetic data. Following the \gls{mbse} methodology, DEVS-BLOOM has been designed with the main objective that any virtual component is built as a \gls{dt} and can be replaced by its real-world counterpart \cite{Marcosig2018}. 

The paper is organized as follows. In the following, we introduce the related work, focused on \glspl{ews}, models of \glspl{hab} behavior, \glspl{usv} trajectory planning, \gls{iot} simulators and all the elements required by the proposed framework. Next, we present the architecture of our framework based on a well-known \gls{ms} formalism. Then we illustrate the simulations performed to test our hypotheses and show the results obtained under different initial conditions. Finally, we draw some conclusions and introduce future lines of research.

%% file: 2_related_work.tex
\section{Related work}\label{sec:related}

As stated above, \glspl{hab} pose severe threats to natural environments. To properly detect, assess, and mitigate these threats in inland waters, it is essential to envision water management from the perspective of an integrative \gls{iot}-based early warning system. \gls{hab}-centric automated \glspl{ews} can effectively help to monitor and treat water bodies since, once deployed, mitigation techniques tailored to those systems can be better designed.

Current \glspl{ews} are supported by a comprehensive set of accurate \textit{base models} that describe the behavior of different elements, such as the dynamics of the water (due to currents and wind) and of the cyanobacteria (due to biological growth, their vertical displacements, and the water dynamics). There exist a significant variability of base models. Eulerian models, for instance, have been used since 1970 to simulate eutrophication, water quality, and biogeochemical processes \cite{Vinccon2019}. These models are composed of differential equations that simulate community dynamics in spaces. Lagrangian models introduce the possibility of adding different classes of particles with individualized properties, although conducting Lagrangian simulations with a large number of particles is a computer-intensive process \cite{VanSebille2018}. Machine learning techniques can also be used to clarify the dynamics of \glspl{hab}. Based on studies from 2008 to 2019, Chen \textit{et al.} show in \cite{Chen2020} numerous applications of machine learning models for predicting various water quality variables, such as salinity, pH, electrical conductivity, dissolved oxygen, ammonium nitrogen, etc. Finally, we may also find mechanistic or process-oriented aquatic models based on knowledge of how target species respond to various ecosystem drivers like nutrient availability, thermal stratification, life cycle characteristics of species, etc. \cite{Rousso2020}. These models can be more appropriate than statistically based models for future predictions. However, they can be challenging because the incomplete knowledge introduced inside the models forces the incorporation of complex Bayesian networks, adding even more uncertainty to the models.

The previous base models are usually integrated inside more generic software tools with advanced \glspl{gui}. For instance, EEMS \cite{EEMS2022} is a \gls{gui} that provides a broad range of pre-processing and post-processing tools to assist in developing, calibrating, and analyzing hydrodynamic, sediment-contaminant, and eutrophication models. MIKE Powered by DHI is a range of software products that enable us to accurately analyze, model and simulate any type of challenge in water environments \cite{MIKE2022}. Delft3D is a set of open source software tools that facilitates modeling subsystems like the hydrodynamic, morphodynamic, waves, water quality, or particle-based subsystems \cite{Delft2022}.

Finally, the aforementioned base models along with the \glspl{gui} are used in \glspl{ews} as forecasting tools \cite{Velez2014,Silva2016}, helping \gls{gcs} operators to make critical decisions. An example close to the authors of this paper is the Spanish Automatic Water Quality Information System, which is a network of nearly 200 automatic alert stations deployed in critical locations of the Spanish hydrographic system to (i) obtain frequent measurements of representative parameters such as water temperature, pH and dissolved oxygen; (ii) provide valuable information about the general quality of the water; and (iii) alert in real time about pollution episodes \cite{Serramia2004}. More examples of \glspl{ews} can be found in other places and settings. The Southeast Environmental Research Center Water Quality Monitoring Network, property of Florida International University, focuses on coastal monitoring of the southern tip of the Florida peninsula and includes some automatic measuring stations that are rotated between the different sampling sites \cite{SERC2022}. United States Geological Survey's National Water Quality Monitoring Network combines data sources and techniques from 110 sites to monitor the U.S. inland waters \cite{Riskin2021}. Environment and Climate Change Canada, in collaboration with the provincial and territorial governments, runs the Freshwater Quality Monitoring and Surveillance program, which encompasses some manual and automatic monitoring networks distributed through the national territory \cite{FWQMS2022}.

The conception, design, and deployment of an \gls{ews} can present complex engineering and systems challenges. To properly monitor and foresee the formation of \glspl{hab}, \glspl{ews} must cover large geographical areas, remain functional over long periods, and include a large variety of sensors, \glspl{usv}, and data in general. Prediction of \glspl{hab} also involves the use of a plethora of base models. A model-driven approach to designing such a complex and heterogeneous infrastructure would help researchers, domain experts, and water authorities to meet design requirements. It also would enable a model-driven control, reducing costs while increasing performance and scalability, and in general, all the benefits derived from applying a \gls{mbse} approach. There exist cases of success in other areas of research like flood detection \cite{Basha2008}, water treatment \cite{Curl2019}, or healthcare \cite{Henares2022}. However, to our knowledge, this is the first research related to developing integrative model-driven solutions for \gls{hab} management. As mentioned above, our approach is integrative because we do not simulate only the water body but also combine the use of \textit{base models} with the help of models of the infrastructure like sensors, \glspl{usv}, \glspl{gcs}, the cloud layer, and even the operator's behavior through a simulation file, which is the main novelty with respect to other approaches in the literature.

%% file: 3_system_architecture.tex
\section{System architecture and design}\label{sec:architecture}

DEVS-BLOOM's model divides the \gls{hab} management system into the three classical \gls{iot} layers: edge, fog, and cloud. The \textit{edge} layer includes all the devices connected to the internet and can generate data. These devices can be sensors, wearables, and other smart devices deployed in the field. The edge layer collects and processes data locally and then sends it to the next layer for further processing. The \textit{fog} layer is an intermediate layer between the edge and the cloud. This layer includes devices with computing power and storage capabilities to perform basic data processing and analysis. The fog layer is responsible for processing data in real time and reducing the amount of data that needs to be sent to the cloud for further processing. The \textit{cloud} layer includes cloud servers and data centers that can store and process large amounts of data. The cloud layer performs complex data analytics and machine learning tasks that require significant computing power and storage capacity \cite {J-Cardenas2021}.

Figure \ref{fig:big_picture} has already illustrated the general picture of the framework architecture. Our \gls{ms} framework is fed with data that may come from the actual water body or from a database, which can, in turn, store authentic or synthetic data. The virtual/real duality developed into some components, modeled as \glspl{dt}, allows DEVS-BLOOM to work in virtual, real, or hybrid modes. The framework works in virtual/simulation mode when data come entirely from the database. Real/controller mode is when data come from the real water body, with actual sensors and \gls{usv} deployed and the system fed with real data. Currently, DEVS-BLOOM is mostly used for infrastructure analysis and design. Thus, data usually come from the database; therefore, DEVS-BLOOM works in virtual/simulation mode. However, sometimes a prototype sensor or \gls{usv} is tested for validation on the field, and then DEVS-BLOOM works in hybrid mode, where some virtual components are simulated and the actual ones are being controlled.

To clarify the specifics of the \gls{devs} nomenclature, we first describe the basic principles of the formalism. Next, the DEVS-BLOOM system architecture is explained.

\begin{figure*}
    \centering
    \includegraphics[width=0.845\textwidth]{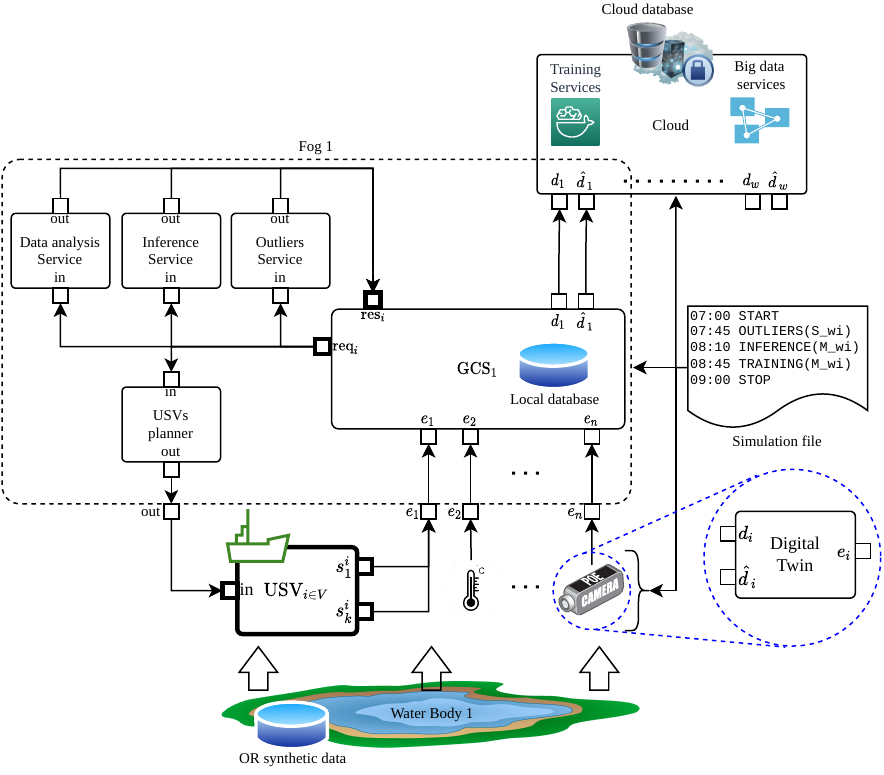}
    \caption{DEVS root coupled model of the system architecture.}
    \label{fig:devs_model}
\end{figure*}

\subsection{The DEVS formalism}

Parallel \gls{devs} is a modular and hierarchical formalism for modeling discrete event systems based on set theory~\cite{Zeigler2018}. It includes two types of models, atomic and coupled, that have an interface consisting of input ($X$) and output ($Y$) ports to communicate with other models. Additionally, in atomic models, every model state ($S$) is associated with the time advance function $ta$, which determines the duration in which the state remains unchanged. 

Once the time assigned to the state has passed, an internal transition is triggered and the corresponding function ($\delta_{\rm int}: S \rightarrow S$) is invoked, producing a local state change ($\delta_{\rm int}(s) = s'$). At that time, the results of the model execution are spread through the output ports of the model by activating an output function ($\lambda$).

Furthermore, external input events (received from other models) are collected in the input ports. An external transition function ($\delta_{\rm ext}: S \times e \times X \rightarrow S$) specifies how to react to those inputs, using the current state ($s$), the elapsed time since the last event ($e$) and the input value ($x$) ($\delta_{\rm ext}((s, e), x) = s'$). Parallel \gls{devs} introduces a confluent function ($\delta_{\rm con}((s, ta(s)), x) = s'$), which decides the next state in cases of collision between external and internal transitions. 

Coupled models are the aggregation/composition of two or more models (atomic and/or coupled), connected by explicit couplings. This makes \gls{devs} closed under coupling and allows us to use networks of systems as components in larger coupled models, leading to hierarchical and modular designs. 

Overall, \gls{devs} provides a framework for information modeling that has several advantages in the analysis and design of complex systems: completeness, verifiability, extensibility, and maintainability.

Once a system is described according to \gls{devs} theory, it can be easily implemented using one of the many \gls{devs} \gls{ms} engines available \cite{J-RiscoMartin2022b}.

DEVS-BLOOM is implemented and executed using xDEVS, a cross-platform \gls{devs} simulator. This library includes a set of C, C++, C\#, Go, Java, Python, and Rust repositories that provide equivalent \gls{devs} interfaces. The project's final goal is to elaborate the fastest \gls{devs} simulation interface with the capacity to simulate models in virtual and real-time and to run simulations in sequential (single-threaded), parallel (multi-threaded), and distributed (not shared memory) architectures. In particular, DEVS-BLOOM uses the xDEVS/Python module of the project. As in xDEVS, our framework can use virtual or real-time. It can run sequential or parallel simulations without modifying a single line of code in the underlying simulation model. 

\subsection{DEVS-BLOOM}

The DEVS-BLOOM root coupled model is depicted in Figure \ref{fig:devs_model}. The components included in this coupled model are: sensors and \glspl{usv} at the edge layer, the fog coupled model, and the cloud atomic model. 

There exist one singular atomic model labeled as \textit{Simulation file} in Figure \ref{fig:devs_model}. It is just a source that reads from a text file all the events that will be injected into the simulation process through its output port. Output and explicit connections related to this atomic model are not explicitly represented in Figure \ref{fig:devs_model} for simplicity because this atomic model is connected to all the components of DEVS-BLOOM. Each entry in the simulation file represents an input event composed of: a time mark indicating the virtual instant in which this event will be triggered, the command type associated with the event, and the arguments each command needs. As a result, this file replicates the set of external events that could happen in a real-world scenario. As the excerpt of Figure \ref{fig:devs_model} illustrates, it always begins and ends with the triggering of the initialization and finalization of the simulation experiment (see START and STOP commands). Some services can be triggered in the middle, like outliers detection or \gls{hab} prediction. The simulation file is a pure virtual element, which does not have an exact match in the real world. In the following sections, we describe the rest of the components included in DEVS-BLOOM.

\subsubsection{Edge layer}

The atomic models in this layer represent edge devices such as environmental sensors, cameras placed at stationary positions, and \glspl{usv}. Particularly, sensors are implemented as \glspl{dt} and can process data from the actual sensor or the database mentioned above. A representation of an atomic sensor model is illustrated in Figure \ref{fig:devs_model}, labeled as \textit{Digital Twin}. Data from its real counterpart is received through the $d_i$ input port. In this case, data is just propagated without extra delays to the corresponding output port $e_i$. On the other hand, data from the database is received by the $\hat{d}_i$ input port. Here the virtual sensor imitates the behavior of the actual sensor, introducing corresponding delays, noise, saturation errors, aging, etc. All these optional parameters are defined through a configuration file. Like most DEVS-BLOOM components, this is a passive atomic model, which is awakened when it receives a \texttt{START} event from the simulation file.

Each \gls{dt} transmits, at their discretion, events that follow a predefined and generic structure that encapsulates the measurements, commands, or any other relevant information. That generic event structure, represented in Figure~\ref{fig:events}, carries a \texttt{timestamp} with the actual event time, a \texttt{source} and \texttt{id} that respectively identify the source and the cause of the event, and a payload which contains a set of key-value pairs with the actual measurements (e.g. \texttt{'Lat': 47.0, 'Lon': -122.0, 'Depth':-0.2, 'TEM': 19.0}). Finally, any time an event is generated, it is transmitted through the corresponding output port $e_i$, which in this case is connected to the fog coupled model of the water body, where the data will be curated and stored in the local fog database.

\begin{figure*}
    \centering
\begin{minipage}[t]{0.30\textwidth}
    \begin{lstlisting}[basicstyle={\ttfamily\scriptsize},escapeinside={<}{>}]

 <id:\textcolor{teal}{'DOX'},>
 <source: \textcolor{red}{'SimSenO'},>
 <timestamp:\textcolor{violet}{'2021-08-23 00:30:05'},>
 <payload:>
 <\textcolor{blue}{\{'Lat':47.50502,'Lon':-122.21501,}>
 <\textcolor{blue}{'Depth':0.00,'DOX':8.69\}}>

    \end{lstlisting}
\end{minipage}
\begin{minipage}[t]{0.65\textwidth}
    \begin{lstlisting}[basicstyle={\ttfamily\scriptsize},escapeinside={<}{>}]
...
<\textcolor{teal}{DOX},\textcolor{red}{SimSenO},\textcolor{violet}{2008-08-23 00:00:00},\textcolor{blue}{\{'id':'DOX','description':'Oxigen Sensor(mg/L)','delay':5,\\
                                                                                                           'max':30.0,'min':0.0,'precision':0.1,'noisesigma':0.2\}}>
...
<\textcolor{teal}{DOX},\textcolor{red}{SimSenO},\textcolor{violet}{2008-08-23 01:30:05},\textcolor{blue}{\{'Lat':47.5050,'Lon':-122.2150,'Depth':0.0,'DOX':11.8\}}>
<\textcolor{teal}{NOX},\textcolor{red}{SimSenN},\textcolor{violet}{2008-08-23 01:30:06},\textcolor{blue}{\{'Lat':47.5050,'Lon':-122.2150,'Depth':0.0,'NOX':0.152\}}>
<\textcolor{teal}{TEM},\textcolor{red}{SimSenT},\textcolor{violet}{2008-08-23 01:30:07},\textcolor{blue}{\{'Lat':47.5050,'Lon':-122.2150,'Depth':0.0
,'TEM':19.6\}}>
...
<\textcolor{teal}{IRA},\textcolor{red}{VirSenI},\textcolor{violet}{2008-08-23 08:30:02},	\textcolor{blue}{\{'Lat':47.5000,'Lon':-122.2200,'IRA': 0.60\}}>
...

\end{lstlisting}
\end{minipage}
    \caption{Generic event structure. On the left, the serialization of a typical event object generated at the edge layer. On the right, an excerpt of messages stored at the local database of fog layer.}
    \label{fig:events}
\end{figure*}

DEVS-BLOOM can handle a set $\mathbf{V}$ of \glspl{usv} atomic models, illustrated as a box with bold lines and ports in Figure \ref{fig:devs_model}. \glspl{usv} receive communications from the fog coupled model or other \glspl{usv} since they are also interconnected to avoid collisions or to share information. Each \gls{usv} has $K$ output ports ($s^{i \in V}_{k \in K}$) to transmit the information gathered by \gls{usv} sensors related to aquatic variables (e.g. temperature, pH, and chlorophyll-a and phycocyanin fluorescence) and to the \gls{usv} state (e.g. power, position).

\subsubsection{Fog layer}

The fog layer is modeled through the \textit{fog} coupled model, which mainly represents the \gls{gcs} associated with the water body. Here, operators and domain experts analyze data, make decisions, and take action. It is worthwhile to mention that DEVS-BLOOM can predict the bloom appearance, automatically guide \glspl{usv} to the zone of interest, or take measurements. Still, all these actions must be validated or complemented by the operators. There can be as many fog-coupled models as water bodies being analyzed by the same cloud infrastructure. Figure \ref{fig:devs_model} represents the first of them. As the Figure shows, the fog coupled model has several input ports that receive the events sent by the \glspl{dt} located at the edge layer (sensors and \glspl{usv}). It also has two output ports that send raw data collected by the sensors to the cloud and augmented or fixed sensor data using outliers detection or data analysis services, through $d_1$ and $\hat{d}_1$ ports, respectively. To reduce visual clutter, Figure \ref{fig:devs_model} does not explicitly represent the coupling relations between fog and cloud. It is quite redundant and makes the Figure unnecessarily large. Basically, $d_1$ and $\hat{d}_1$ are connected through two additional external output couplings (from GCS$_1$ to Fog$_1$) and two internal couplings (from Fog$_1$ to Cloud). The fog coupled model contains several atomic models, detailed below.

The \textit{\gls{gcs} atomic model} represents the core of the computing infrastructure of the control station. It is usually a static workstation or laptop connected to the local network. This simplified \gls{dt} receives simulation commands from the simulation file atomic model, which tell the computer when to start reading data, execute an outliers detection service, an inference over the \gls{hab} predictive models, \glspl{usv} path planning, etc. When the simulation starts, sensor data are received through the $e_i$ input ports and stored in the local database. These data are sent through the $d_1$ fog output port, which is connected to the $d_1$ cloud input port. On the other hand, when a service request is received from the simulation file, it is propagated through the output port $req_i$, which is connected to the corresponding atomic model. This port is drawn in bold in Figure \ref{fig:devs_model} because it represents a set of output ports. Fixed or predicted data are also stored in the local database and regularly sent through the $\hat{d}_1$ output port, connected to the $\hat{d}_1$ cloud input port.

The fog coupled model also has a set of atomic models in charge of executing services. They are currently part of the \gls{gcs}$_1$ atomic model in the real system. Still, we have decided to implement them as external atomic models to separate the services, models, or functions that they incorporate. These atomic models receive commands from the \textit{in} input port and send the results through the \textit{out} output ports. These output ports are connected back to the \gls{gcs} or the \gls{usv} atomic models, controlling the navigation system of the \glspl{usv}. We have currently deployed four services: one to detect and fix outliers, labeled as \textit{Outliers services} in Figure \ref{fig:devs_model}, another one to perform inference and compute the probability of \gls{hab} formation and location in the water body, labeled as \textit{Inference service}, a third one to carry out data analysis over the database and generate reports, named \textit{Data analysis service}, and the last one is the \glspl{usv} path planner, as labeled in Figure \ref{fig:devs_model}, which taking the probabilities computed by the inference service calculates and sends the waypoints and trajectories that \glspl{usv} must follow.



\subsubsection{Cloud layer}

Finally, the \textit{cloud atomic model} is located in the cloud layer. It receives all the data from different water bodies (raw and estimated, i.e., fixed or predicted) and stores them in the central cloud database. As in the fog coupled model, the cloud atomic model can run different services but is highly scaled to handle one or several water bodies. These services include executing big data analyses involving all the data stored in the central database or running training services to update current inference models located at the fog-coupled models. In any case, these actions are always triggered by the simulation file. We have not included dedicated atomic models to run services because they are always processes installed in docker containers, i.e., they have a distributed architecture. They do not need to be encapsulated as \gls{devs} models, i.e., the cloud layer is viewed as a centralized entity.



DEVS-BLOOM scalability is ensured since our framework has been designed to consider possible scalability issues, allowing users to externalize and scale critical services in case of a bottleneck. For instance, \gls{devs} atomic models that consume more resources, like those designed to run training or data analysis services, can be easily simulated in parallel or distributed computing architectures \cite{J-RiscoMartin2022}.

%% file: 4_experiments.tex
\section{Use cases}\label{sec:experiments}

This Section presents two simulation scenarios to test the developed framework. Additionally, this Section illustrates how an actual sensor in DEVS-BLOOM can replace a virtual one and show some advances in the design of the real \gls{usv}, which is evolving with the corresponding DEVS-BLOOM \gls{usv} model.

\begin{figure*}[ht]
  \centering
  \includegraphics[width=0.8\textwidth]{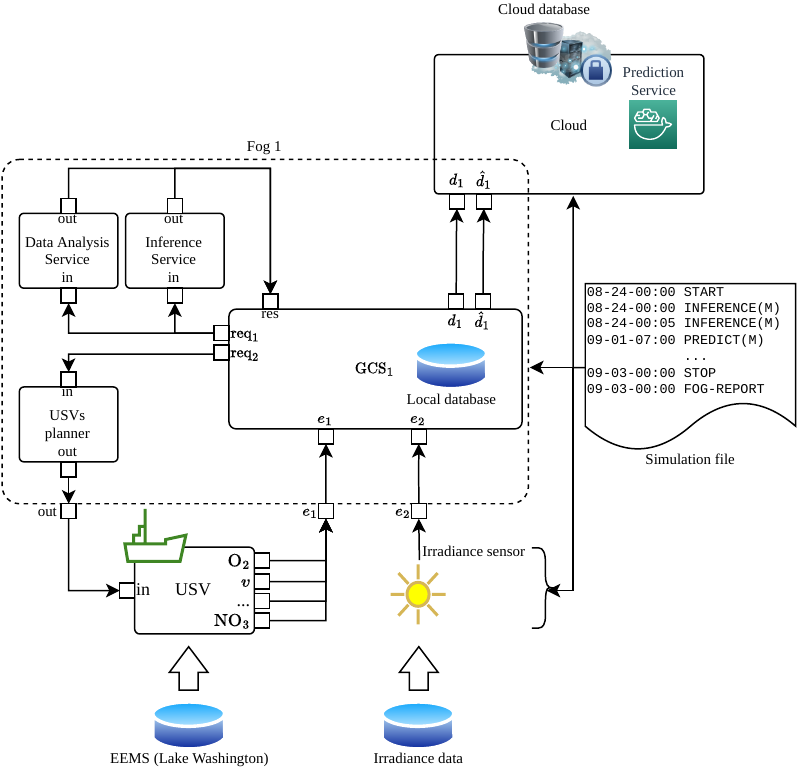}
  \caption{DEVS-BLOOM root coupled model of the use case.}
  \label{fig:devs_use_case}
\end{figure*}

Figure \ref{fig:devs_use_case} shows the DEVS-BLOOM model used in this Section. It is a particular case of the general model depicted in Figure \ref{fig:devs_model}, used to monitor a water body corresponding to an area of Lake Washington.

We provide more details of each atomic model instance included in Figure \ref{fig:devs_use_case} throughout each use case.

\subsection{Monitoring use case}

The monitoring scenario is relevant for operators and domain experts in charge of the \gls{gcs} and local operative decisions, monitoring \glspl{hab} state and evolution through the use of a \gls{usv}, i.e., it shows how DEVS-BLOOMS is used to predict the next location of the \gls{hab} and to automatically control the \gls{usv} to follow the position and confirm the prediction.

In this case, the whole water body dataset is synthetic and generated with the EEMS tool \cite{EEMS2022}, which incorporates an accurate model of Lake Washington. It allows us the artificial generation of \glspl{hab}. As a result, DEVS-BLOOM receives EEMS input data (see Figure \ref{fig:devs_use_case}) that includes water speed, water temperature, oxygen and nitrates densities, and for validation of our framework, algae concentration. 

Additionally, as Figure \ref{fig:devs_use_case} shows, we have included a virtual irradiance sensor, which generates synthetic irradiance data taken from PVGIS\footnote{https://re.jrc.ec.europa.eu/pvg\_tools}. Neither \gls{eems} nor PVGIS give stochastic data, so there is no need to proceed with Monte Carlo simulations.

Our scenario has at the edge layer a \gls{usv} that must monitor the water and transmit data to the fog and cloud layers. As Figure \ref{fig:devs_use_case} depicts, the \gls{usv} is instrumented with several sensors and units. Some of them take data from the water body to continuously monitor the state of the bloom and feed the inference model, and others from internal component models:

\begin{itemize}
\item Temperature sensor: is in charge of measuring the water temperature. This signal influences the calibration of other sensors and the growth dynamics of the bloom.
\item Power unit: includes solar panels, chargers, and batteries in charge of recharging the boat's batteries when it receives solar radiation. For this scenario, we have included the following base model:
\begin{eqnarray}
prop & = & K_p \cdot \sqrt{e_{lat}^2 + e_{lon}^2} \nonumber \\
power & = & K_e + K_s \cdot sun - prop \nonumber 
\end{eqnarray}
$K_p=30$ is the propulsion constant, $K_e=-0.003$ represent the electronic power consumption, $K_s=0.04$ is the sun power factor, $prop$ is the resultant propulsion, $e_{lat}$ and $e_{lon}$ are the latitude and longitude error of the \gls{usv} with respect to the \gls{hab} position, computed by the \gls{usv} planner atomic model, $power$ is the battery energy level, and $sun$ is the normalized irradiance value.
\item Flow meter: measures the speed and direction of the water with respect to the ship. We may infer the water's speed and direction by discounting the ship's speed.
\item Positioning unit: allows us to measure the position and speed of the ship, following these two equations:
\begin{eqnarray}
lat_{usv} & = & e_{lat} + K_{2d} \cdot wfv \nonumber \\
lon_{usv} & = & e_{lon} + K_{2d} \cdot wfu \nonumber
\end{eqnarray}
$K_{2d}=0.01$ is the 2D \gls{usv} displacement constant, and $(wfv, wfu)$ is the water speed (north and east components).
\item Dissolved oxygen probe: is in charge of measuring the dissolved oxygen density in the water. If there are high levels of oxygen, there may be a bloom of algae that produces oxygen by photosynthesis.
\item Nitrogen probe: measures the density of dissolved nitrates in the water. Nitrate is the main food for algae. Therefore, the inference service uses this signal to predict the bloom's growth.
\end{itemize}

During the simulation, irradiance and \gls{usv} sensors capture measurements and send them to the fog layer. We utilize the inference service in this layer, shown in Figure \ref{fig:devs_use_case}. It has a predictive model based on differential equations that, using water speed,  temperature, coordinates, oxygen and nitrates densities, and solar irradiance, anticipates the emergence and displacement of \glspl{hab} as follows: 

\begin{eqnarray}
\frac{dr(t)}{dt} & = & K_1 \cdot photo(t) + K_2 \cdot breath(t) \nonumber \\
                 &   & - K_3 \cdot (r(t) - r(0)) \nonumber \\              
\frac{dlat_{bloom}(t)}{dt} & = & K_v \cdot wfv(t) \nonumber \\
\frac{dlon_{bloom}(t)}{dt} & = & K_v \cdot wfu(t) \nonumber \\
photo(t) & = & sun(t) \cdot nox(t) \nonumber \\
breath(t) & = & dox(t) \cdot nox(t) \nonumber
\end{eqnarray}

In the previous equation, $r$ represents the bloom density, while $photo$ and $breath$ represents photosynthesis and respiration, respectively. Besides, $(lat, lon)$ are the coordinates (latitude and longitude) of the position of the bloom at a given height, whereas $(wfv,wfu)$ is the water velocity at the same coordinates. $nox$ and $dox$ are nitrogen and oxygen concentration, respectively (mg/l). Regarding the constants, $K_1=5.0$ and $K_2=0.05$ represent the \gls{hab} growth constant, whereas $K_3=0.17$ is the decay constant. $K_v=0.0167$ represents the percentage of the water velocity transferred to the \gls{hab}. The values of the constants are initially obtained by training the system with the least squares method.

Then the \glspl{usv} planner in Figure \ref{fig:devs_use_case} generates track points for the \gls{usv}. In this preliminary version, the planner computes the error between \gls{usv} and \gls{hab} positions as follows:

\begin{eqnarray}
e_{lat} & = & lat_{bloom} - lat_{usv} \nonumber \\
e_{lon} & = & lon_{bloom} - lon_{usv} \nonumber
\end{eqnarray}

To close the loop, the \gls{usv} navigates to the track point and retakes measurements. During the simulation, all the data is saved into the fog and cloud databases, which can be plotted and analyzed in real time. The Data Analysis Service depicted in Figure \ref{fig:devs_use_case} can be activated to automate this process. This atomic model executes a set of functions to create all the figures and videos of interest for the operator or the domain expert. Details about implementing these automatically generated reports can be found in \cite{HerguedasPinedo2023}.

\begin{figure*}[ht]
    \centering
    \includegraphics[width=0.45\textwidth]{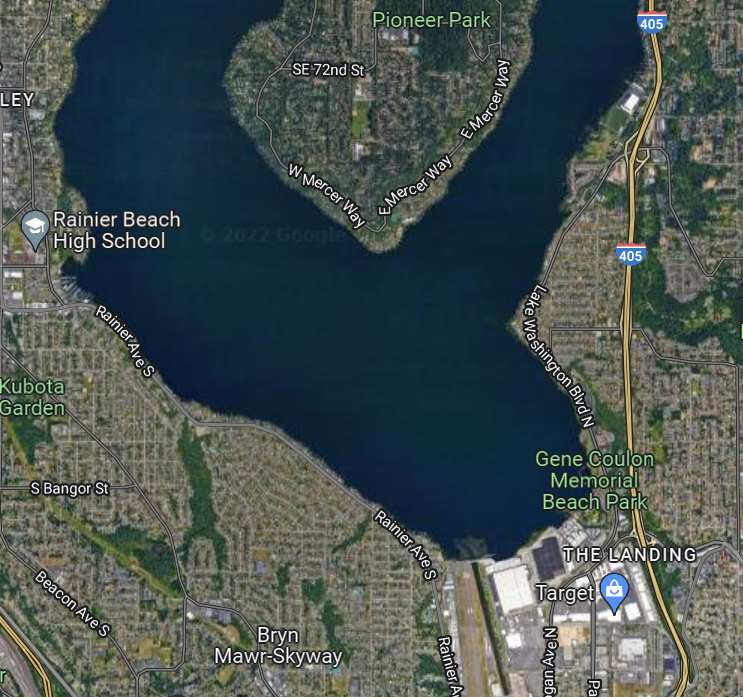}
    \caption{Lake Washington area.}
    \label{fig:sim0}
\end{figure*}

In the following, we show the simulation results. Figure \ref{fig:sim0} shows the lake area where \glspl{hab} are forming. The lower part of the image shows how a channel flows into the lake in a shallow area. Such areas are known as incubators because they provide ideal conditions for forming blooms and accumulations of nitrates in areas with solar radiation. The inference model is initialized near the incubator at the beginning of the day. It is very likely that the bloom is born in this area, then grows with solar radiation, moves with the water currents, and disperses throughout the rest of the lake.

Figure \ref{fig:photo} illustrates the simulation state while tracking a \gls{hab}. As mentioned above and depicted at the bottom of Figure \ref{fig:devs_use_case}, at this stage of the project, all the measured data from the water body are from \gls{eems}, except for the irradiance values that are taken from PVGIS since \gls{eems} does not include these. The rest of the data (\glspl{usv} battery status, bloom displacement prediction, etc.) come from our models. Next, we describe each plot in Figure \ref{fig:photo}:

\begin{figure*}[ht]
    \centering
    \includegraphics[width=1\textwidth]{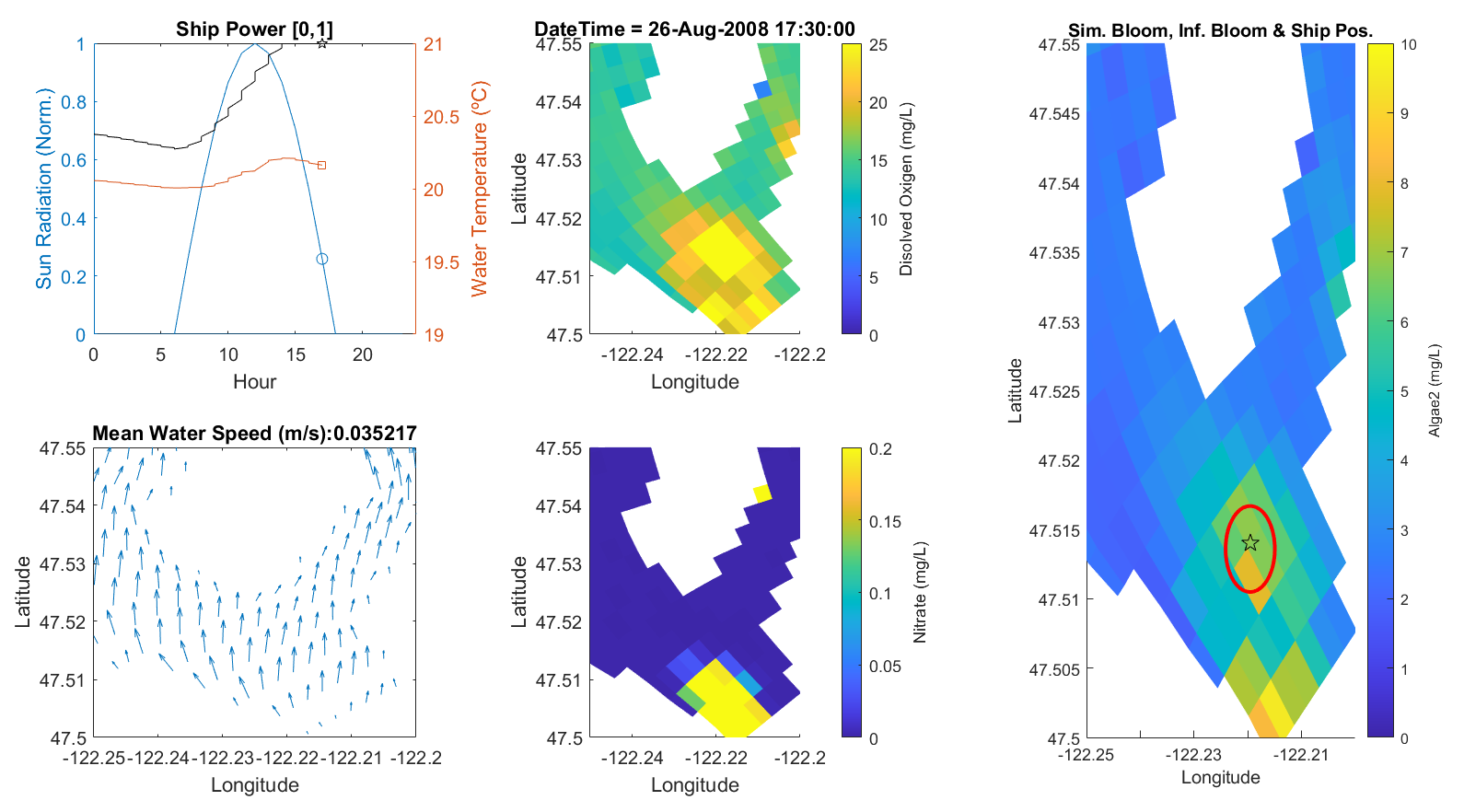}
    \caption{Frame of bloom tracking simulation: (upper-left) USV measured signals , water temperature, and solar irradiance, (lower-left) water speed. (top-center) oxygen density, (bottom-middle) nitrate density, (right) \gls{hab} \gls{eems} given density for validation, \gls{hab} prediction as a red circle and ship position as a star.}
    \label{fig:photo}
\end{figure*}

\begin{itemize}
\item The upper left graph shows the signals measured by the \gls{usv} and the irradiance sensor as a function of the time of day: sun radiation (blue), water Temperature (red), and ship's electric power (black). At the time of the simulation, Figure \ref{fig:photo} shows that the solar panels have fully charged the ship batteries.
\item The lower left graph shows the map of water direction and velocity in the surface layer. The ship measures this signal at its position and reports it to the fog layer to estimate the bloom displacement. The simulator also uses the information from this map to perturb the ship dynamics.  
\item The top center graph shows the map of the dissolved oxygen density in the surface layer. The \gls{usv} takes this measurement, and the inference model uses it to decide whether there is a bloom or not.  
\item The bottom middle graph shows the map of nitrate density on the surface. The inference model takes this measurement obtained by the \gls{usv} to estimate the bloom growth. 
\item The right graph shows the \gls{hab} density map in the surface layer, the inferred bloom (red circle), and the \gls{usv} position. The \gls{hab} density map is data directly taken from \gls{eems} to validate that the inference model is correctly predicting the \gls{hab} dynamic.
\end{itemize}

The full simulation video can be found in \cite{Video2022}.

\begin{figure*}
    \centering
    \includegraphics[width=0.7\textwidth]{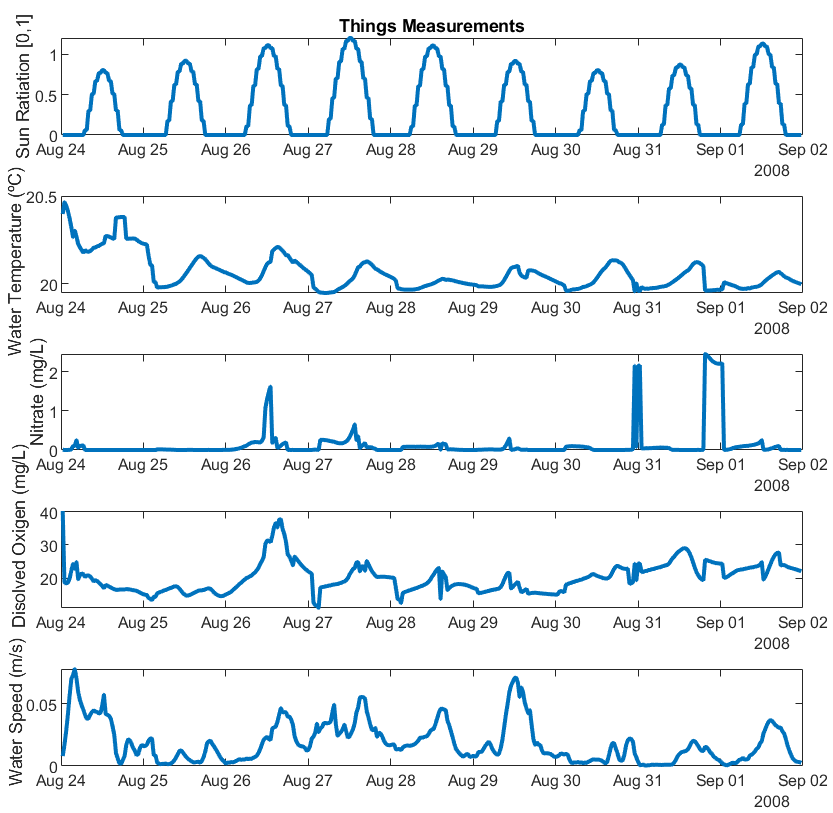}
    \caption{Sensors' signals.}
    \label{fig:sensors}
\end{figure*}

As mentioned above, all the data used in this simulation are synthetic. Consequently, all the sensors work on virtual mode, as \glspl{dt}. When a sensor must take a measurement, it searches the database (the EEMS file or the irradiance database), modifies the signal according to its technical characteristics, and generates a message with the signal value. The fog layer receives these signals to perform different calculations like the model inference and periodically uploads them to the cloud layer. Figure \ref{fig:sensors} shows the signal values recorded by all the sensors of this use case after several (virtual) days of simulation.

\begin{figure*}
    \centering
    \includegraphics[width=0.7\textwidth]{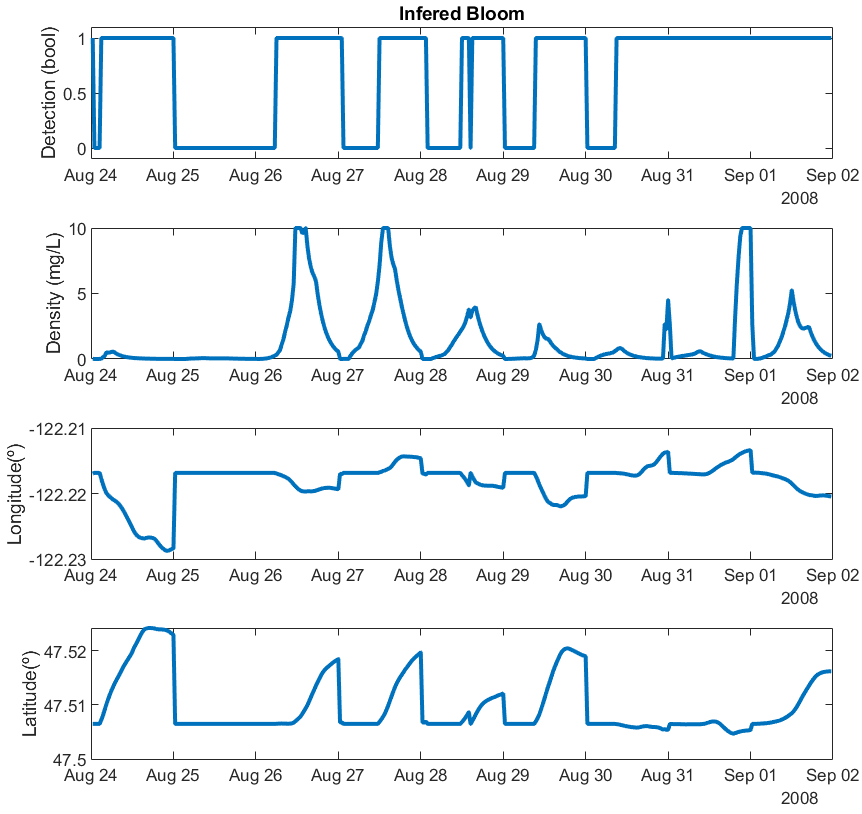}
    \caption{Bloom Inference model.}
    \label{fig:inference}
\end{figure*}

Figure \ref{fig:inference} shows the evolution of the \gls{hab} inference model. The first plot shows a boolean value indicating whether the bloom has been detected or not. The second plot shows the estimated bloom density. The third and fourth plots show the displacement estimation: longitude and latitude. Figure \ref{fig:inference} shows how blooms are detected and monitored almost every day. Some of these blooms have significant densities and move around significantly, requiring dynamic monitoring.

\begin{figure*}
    \centering
    \includegraphics[width=0.7\textwidth]{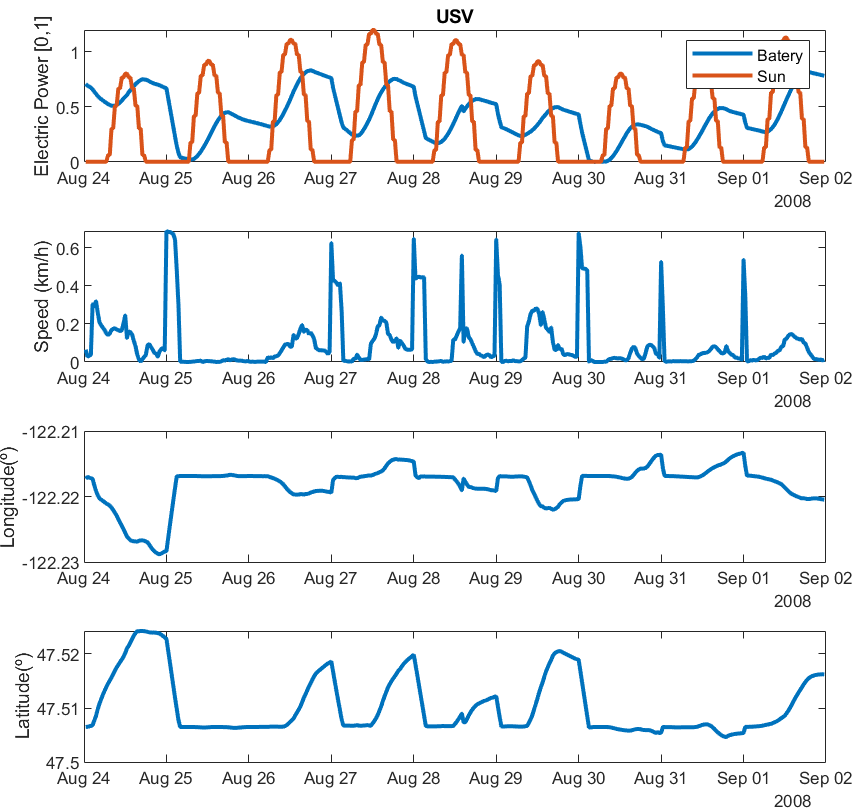}
    \caption{\gls{usv} model.}
    \label{fig:usv}
\end{figure*}

Finally, Figure \ref{fig:usv} depicts the status of the \gls{usv} model. The first graph shows the status of the power unit. The second plot shows the velocity of the \gls{usv}. The third and fourth graphs show the position, longitude, and latitude. On August 30, the Figure shows that the \gls{usv} runs out of battery since it has been tracking blooms to distant points for four consecutive days. 

\subsection{Prediction use case}

The second use case is relevant for water authorities. It consists of predicting \glspl{hab} in the coming days based on weather forecasts. At the end of the day, the \gls{gcs} in Figure \ref{fig:devs_use_case} uploads all this information to the cloud layer. All the data history is available in this layer, allowing us to use the predictive model to analyze medium or long-term events.

To predict future blooms, a \emph{Prediction Service} atomic model has been implemented in the cloud layer. This service is responsible for predicting the occurrence of upcoming \glspl{hab} and their evolution from weather forecasts. These predictions depend highly dependent on local conditions, so they must be designed ad hoc. In our case, in this area of the lake, there is a source of nitrates or dissolved sediments, which is activated by rainfall. At ideal water temperatures, these dissolved sediments and the sunlight are the main precursors of \glspl{hab}. From these precursors, bloom growth can be predicted. On the other hand, surface water currents can be inferred from wind forecasts, which can be used to predict the \gls{hab} displacement.

\begin{figure*}
    \centering
    \includegraphics[width=0.7\textwidth]{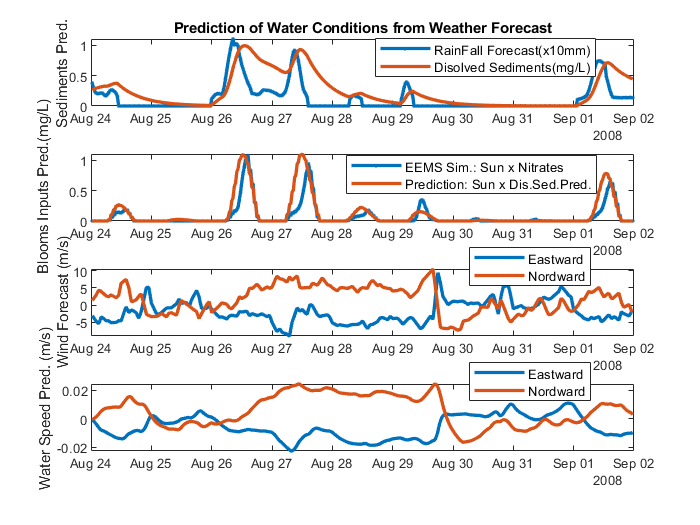}
    \caption{Water and dissolved sediments state inferreed from wind, sun and rainfall forecasts.}
    \label{fig:wf}
\end{figure*}

Firstly, the state of water and dissolved sediments are inferred from wind, sun, and rainfall forecasts. Figure \ref{fig:wf} shows the results of this inference, comparing it with the results generated with \gls{eems}. The first plot shows the rainfall forecast and the inference of dissolved sediments, which follows a simple exponential model. The second plot shows the bloom precursor signal, Sun-Nitrates, the values generated by \gls{eems} and those inferred by the service. The third plot shows the wind forecast, and the fourth plot shows the inferred values for the water speed.

\begin{figure*}
    \centering
    \includegraphics[width=0.7\textwidth]{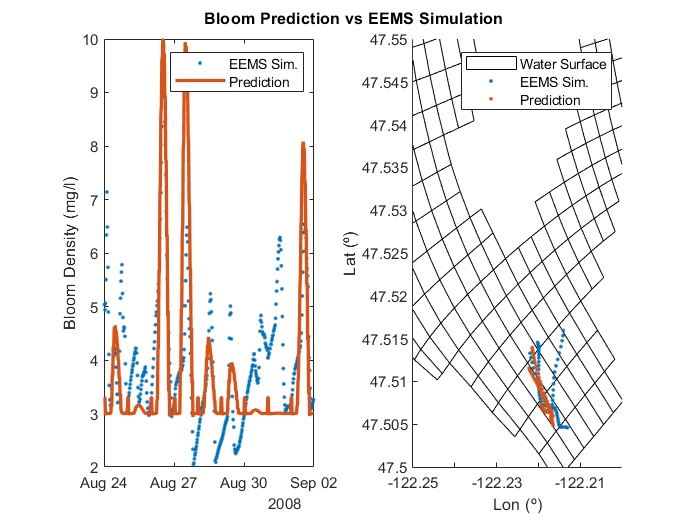}
    \caption{Bloom prediction.}
    \label{fig:bf}
\end{figure*}

Next, the \emph{Prediction Service} atomic model computes the \gls{hab} state from the previous results. Figure \ref{fig:bf} shows the final output, comparing it to the results simulated with \gls{eems}. The plot on the left shows the \gls{hab} density generated by \gls{eems} versus the density predicted by the atomic model. It can be seen that it correctly predicts the 60\% of the bloom cases. The graph on the right shows the trajectory of these \glspl{hab}, predicting where the bloom will move accurately in most cases.

\subsection{Integration of real sensors and \gls{usv} design}

DEVS-BLOOM uses the xDEVS/Python library. xDEVS/Python can simulate models in real-time \cite{J-RiscoMartin2022b}. A scaling factor can be provided, transforming hours into minutes, minutes into seconds, etc. This is important when incorporating hardware in the loop to the virtual framework\cite{Niyonkuru2021} since, for instance, the previous use case handles periods of 30 minutes, but we may want to perform tests with sensors sending data every minute. Additionally, xDEVS can interrupt the real-time simulation with the arrival of  data sent by an external hardware device. To do this, the root coupled model must have an input port to inject data, and an atomic model must handle the arrival of this data through its external transition function.

\begin{figure*}
    \centering
    \begin{subfigure}[b]{0.47\textwidth}
      \includegraphics[width=\textwidth]{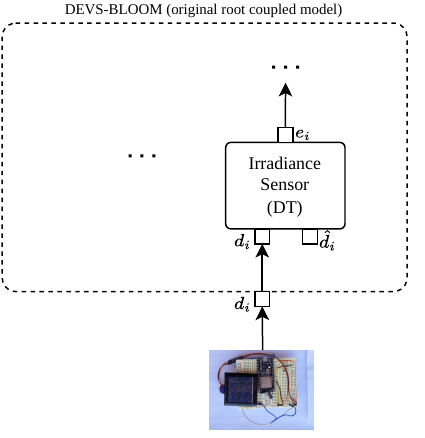}
      \caption{Integration scheme of an actual sensor in DEVS-BLOOM.}
      \label{fig:devs-hil-1}
    \end{subfigure}\hfill
	\begin{subfigure}[b]{0.47\textwidth}
      \includegraphics[width=\textwidth]{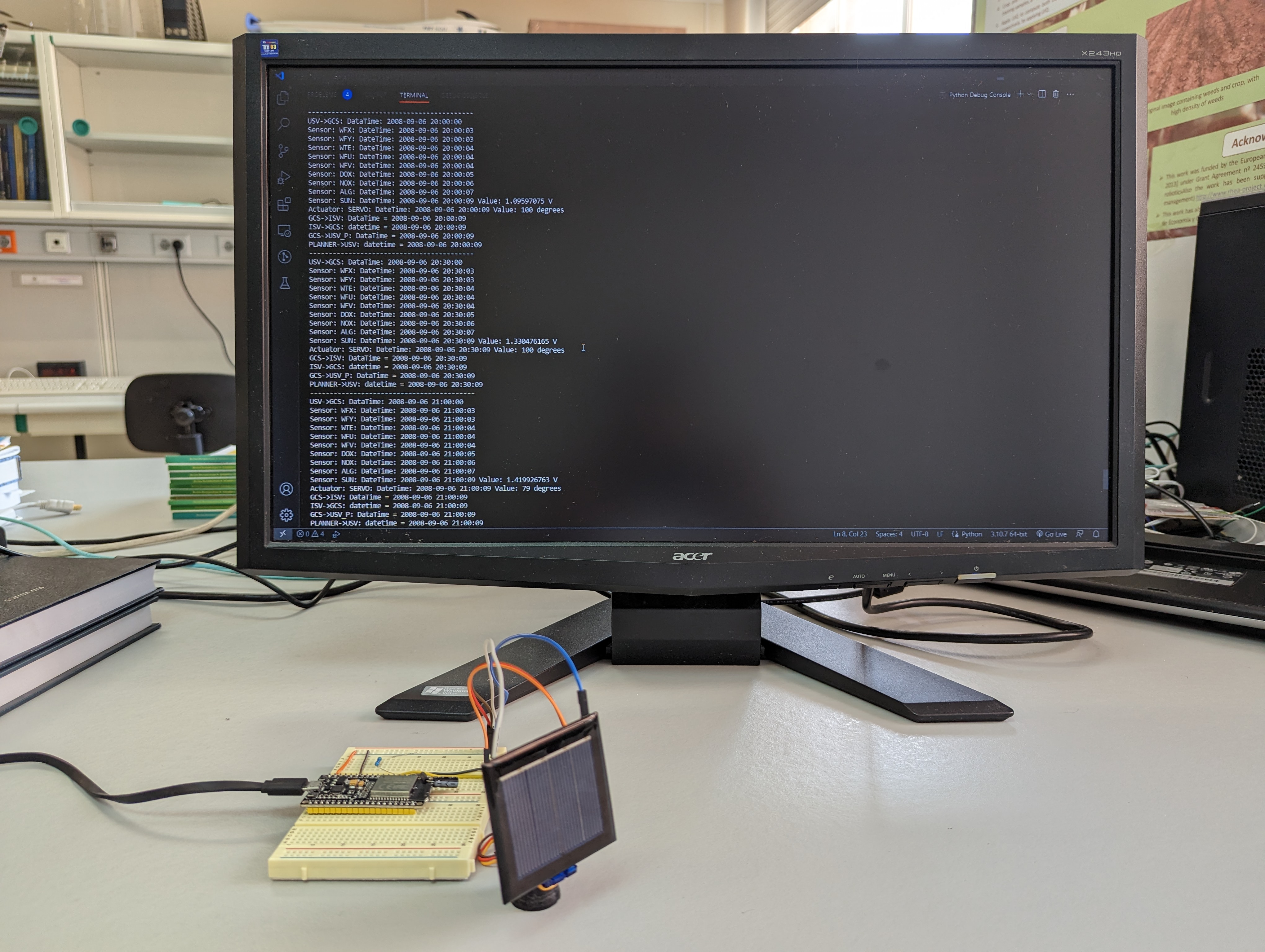}
      \caption{DEVS-BLOOM real-time simulation with the actual sensor.}
      \label{fig:devs-hil-2}
    \end{subfigure}
	\begin{subfigure}[b]{0.47\textwidth}
      \includegraphics[width=\textwidth]{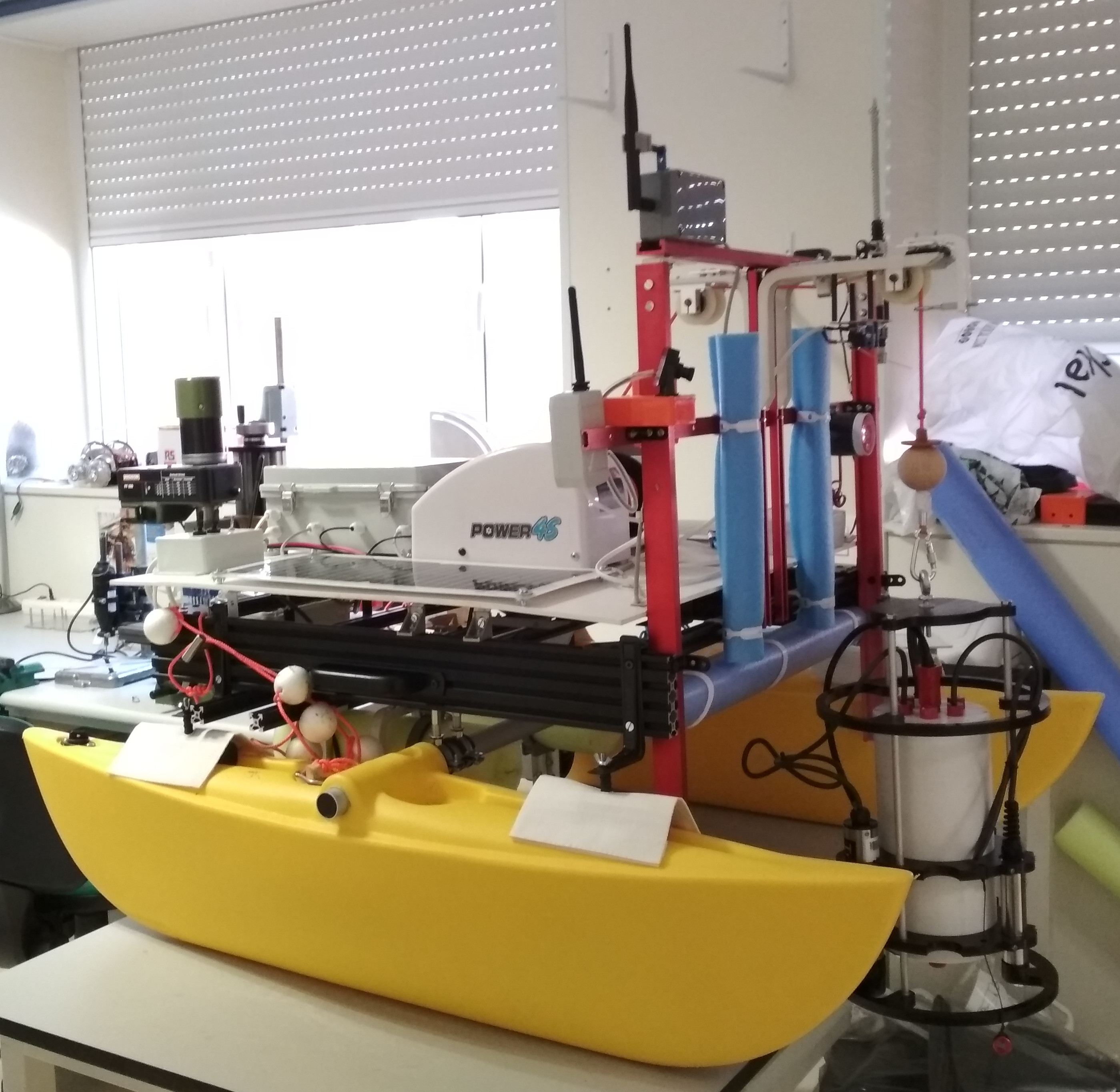}
      \caption{\gls{usv} first prototype.}
      \label{fig:devs-hil-3}
    \end{subfigure}\hfill
    \begin{subfigure}[b]{0.47\textwidth}
      \includegraphics[width=\textwidth]{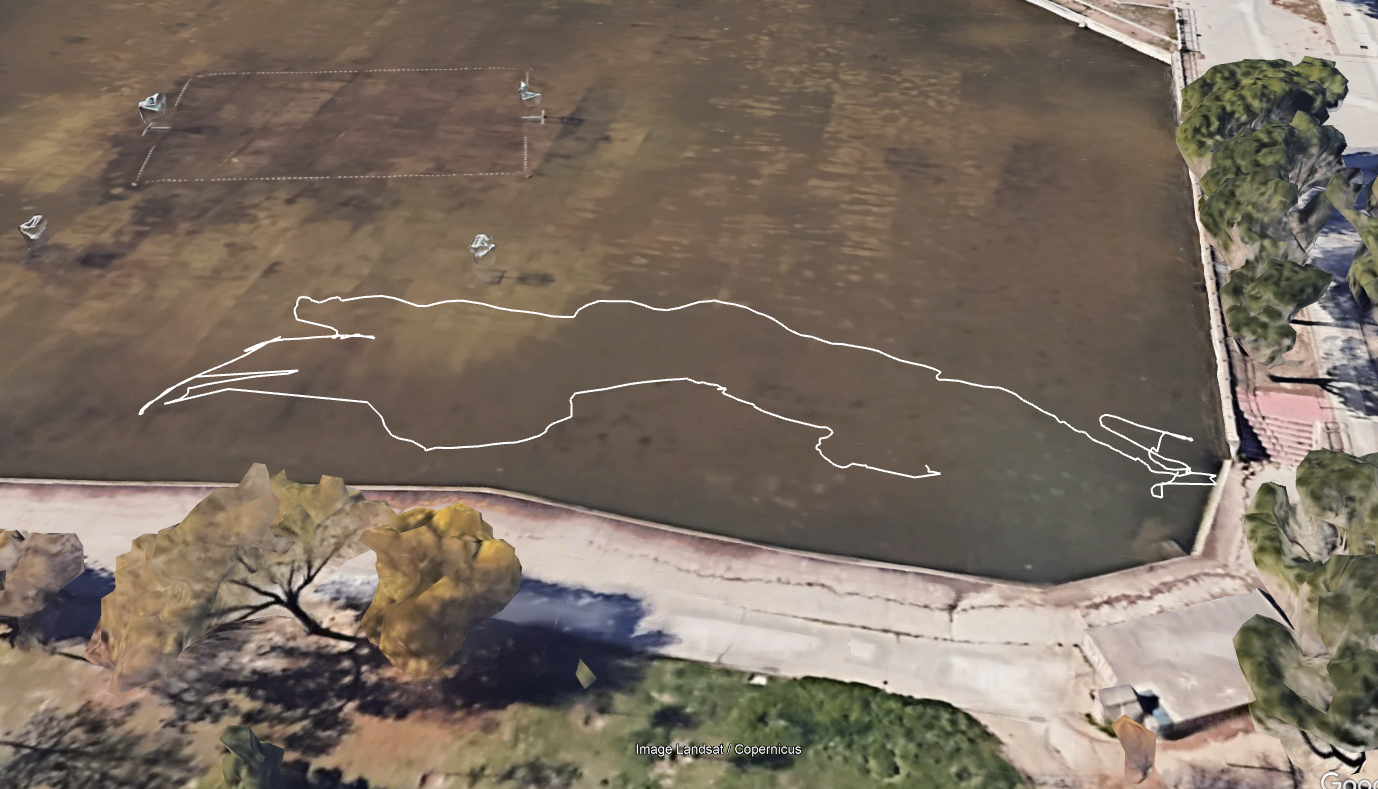}
      \caption{Trajectory registered by the \gls{usv} in a navigation test.}
      \label{fig:devs-hil-4}
    \end{subfigure}
    \caption[HIL]{Integration of real sensors and \gls{usv} design.}
    \label{fig:devs-hil}
\end{figure*}

To demonstrate the ability of DEVS-BLOOM to integrate actual sensors, we have used the xDEVS characteristics mentioned above with the irradiance sensor. Figure \ref{fig:devs-hil-1} depicts schematically how the real sensor is connected to the original atomic model shown in Figure \ref{fig:devs_use_case}. To this end, we use the input port $d_i$ explained in Figure \ref{fig:devs_model}, adding an input $d_i$ port to the root coupled model. xDEVS/Python automatically manages the communication between the sensor and DEVS-BLOOM through a software handler. The procedure is relatively straightforward since the external transition function of the sensor \gls{dt} is automatically triggered when the actual sensor injects data.

On the other hand, Figure \ref{fig:devs-hil-2} shows a picture of a real-time execution, where data received by the actual sensor is correctly logged by DEVS-BLOOM. This procedure also allows us to validate the virtual sensor model, tuning its parameters (delay, precision, noise, etc.) if necessary. The predictive algorithms automatically manage failures in sensors. There is an outliers detection phase before the prediction, where outliers and missing data are replaced by regression. An alarm is triggered in case of failure, and the domain expert can take action if necessary. The parallel \gls{devs} formalism is of great help when dealing with these issues.

New sensors are acquired and tested through our framework as the project evolves. Currently, the most challenging part is the \gls{usv} design. Figure \ref{fig:devs-hil-3} shows our first \gls{usv} prototype with all the sensors embedded, and Figure \ref{fig:devs-hil-4} depicts one of the controlled tests to validate the navigation system. As the \gls{usv} evolves, the DEVS-BLOOM virtual model does the same to match the behavior of the real counterpart \cite{FerreroLosada2023}.

As it can be seen, DEVS-BLOOM can help us to design an integral \gls{ews} considering different elements and exploring all the alternatives. Our \gls{ms} framework facilitates the elaboration of sustainable and efficient \gls{hab} management systems while saving costs with well-dimensioned instruments, \glspl{usv}, and \glspl{gcs}.

%% file: 5_conclusion.tex
\section{Conclusion and future work}\label{sec:conclusion}

\glspl{hab} induce severe threats to water quality. To properly detect, assess, and mitigate these threats to water infrastructures, it is necessary to envision well-structured and robust methods to perform continuous monitoring and to deploy efficient infrastructure and proactive strategies to reduce their adverse effects. \gls{cps} integrative \gls{ms} is crucial to reaching these objectives since it provides sustainable mechanisms to analyze algorithms and the infrastructure we may need to deploy such systems. However, current approaches do not combine the analysis of \textit{base} models and algorithms with the infrastructure.

In this paper, we have introduced DEVS-BLOOM, a novel \gls{ms} framework to enable real-time monitoring and hazard prediction of \glspl{hab} while analyzing the effectiveness of infrastructure deployment. Our framework can automatically manage the design of advanced \glspl{ews} and propose decisions over the evolution of \glspl{hab}. Our approach is based on solid principles of \gls{mbse} and the \gls{devs} \gls{ms} formalism. Furthermore, the entire infrastructure can be modeled upon the \gls{iot} and \gls{dt} paradigms. DEVS-BLOOM allows an incremental design, assuring reliability and scalability to multiple water bodies and minimizing costs in the conception of the final installations. Additionally, all the predictive models designed in the \gls{ms} phase can be later used in the real infrastructure. Our framework also allows different resolution views, for the interpretation of a domain expert at the fog layer and the interpretation of water authorities at the cloud layer, following the \gls{iot} nomenclature. 

Future work includes, on the one hand, the inclusion of new models (e.g., related to the \glspl{usv} dynamics) into DEVS-BLOOM, the improvement of its visualization tools, or the validation of the current \gls{hab} models against a real scenario. On the other hand, we plan to incrementally replace all the elements in the simulated model with those in a real-world use case, complementing the virtual representation of the system introduced in this paper with its final deployment.

Finally, we want to highlight that having a scientific framework to predict \glspl{hab} formation and to take management actions also provides an organizing principle for fundamental research. This framework will serve and benefit the engagement of theory with \gls{ms} foundations. Complementary \gls{hab} research on mathematical models or systems engineering can be easily integrated into our DEVS-BLOOM framework. It will improve the scientific exploitation of discoveries and support the development of new bases for forecasting future effects on water quality and other sustainable water ecological challenges such as wastewater recycling or smart agriculture.

%% file: main.bbl
\begin{thebibliography}{10}
\providecommand{\url}[1]{\texttt{#1}}
\providecommand{\urlprefix}{URL }
\expandafter\ifx\csname urlstyle\endcsname\relax
  \providecommand{\doi}[1]{DOI:\discretionary{}{}{}#1}\else
  \providecommand{\doi}{DOI:\discretionary{}{}{}\begingroup
  \urlstyle{rm}\Url}\fi
\providecommand{\eprint}[2][]{\url{#2}}

\bibitem{Vincent2009}
Vincent WF.
\newblock Cyanobacteria.
\newblock \emph{Encyclopedia of Inland Waters} 2009; 3: 226--232.

\bibitem{Schmale2019}
Schmale DG, Ault AP, Saad W et~al.
\newblock Perspectives on {H}armful {A}lgal {B}looms ({HAB}s) and the
  cyberbiosecurity of freshwater systems.
\newblock \emph{Frontiers in Bioengineering and Biotechnology} 2019; 128.

\bibitem{Meriluoto2017}
Meriluoto J, Spoof L and Codd GA.
\newblock \emph{Handbook of cyanobacterial monitoring and cyanotoxin analysis}.
\newblock John Wiley \& Sons, 2017.

\bibitem{Ung1972}
Ung MT and Moellmer WO.
\newblock A practical example of the use of simulation to develop an optimal
  plan for water—quality control.
\newblock \emph{SIMULATION} 1972; 19(4): 109--117.
\newblock \doi{10.1177/003754977201900402}.
\newblock \urlprefix\url{https://doi.org/10.1177/003754977201900402}.
\newblock \eprint{https://doi.org/10.1177/003754977201900402}.

\bibitem{Long2011}
Long TY, Wu L, Meng GH et~al.
\newblock Numerical simulation for impacts of hydrodynamic conditions on algae
  growth in {C}hongqing section of {J}ialing river, {C}hina.
\newblock \emph{Ecological Modelling} 2011; 222(1): 112--119.
\newblock \doi{10.1016/j.ecolmodel.2010.09.028}.
\newblock
  \urlprefix\url{https://www.sciencedirect.com/science/article/pii/S0304380010004990}.

\bibitem{Pyo2019}
Pyo J, Duan H, Baek S et~al.
\newblock A convolutional neural network regression for quantifying
  cyanobacteria using hyperspectral imagery.
\newblock \emph{Remote Sensing of Environment} 2019; 233: 111350.
\newblock \doi{10.1016/j.rse.2019.111350}.
\newblock
  \urlprefix\url{https://www.sciencedirect.com/science/article/pii/S0034425719303694}.

\bibitem{EEMS2022}
DSI.
\newblock {EE} {M}odeling {S}ystem.
\newblock https://www.eemodelingsystem.com, 2022.
\newblock
  \urlprefix\url{https://www.eemodelingsystem.com/user-center/downloads}.

\bibitem{Wu2022}
Wu Y, Zhang J, Hou Z et~al.
\newblock Seasonal dynamics of algal net primary production in response to
  phosphorus input in a mesotrophic subtropical plateau lake, southwestern
  {C}hina.
\newblock \emph{Water} 2022; 14(5).

\bibitem{Wymore2018}
Wymore AW.
\newblock \emph{Model-based systems engineering}, volume~3.
\newblock CRC press, 2018.

\bibitem{Zeigler2018}
Zeigler BP, Muzy A and Kofman E.
\newblock \emph{Theory of modeling and simulation: discrete event \& iterative
  system computational foundations}.
\newblock Academic press, 2018.

\bibitem{Marcosig2018}
Marcosig EP, Giribet JI and Castro R.
\newblock {DEVS-OVER-ROS (DOVER)}: {A} framework for simulation-driven embedded
  control of robotic systems based on model continuity.
\newblock In \emph{2018 Winter Simulation Conference (WSC)}. pp. 1250--1261.
\newblock \doi{10.1109/WSC.2018.8632504}.

\bibitem{Vinccon2019}
Vin{\c{c}}on-Leite B and Casenave C.
\newblock Modelling eutrophication in lake ecosystems: a review.
\newblock \emph{Science of the Total Environment} 2019; 651: 2985--3001.

\bibitem{VanSebille2018}
Van~Sebille E, Griffies SM, Abernathey R et~al.
\newblock Lagrangian ocean analysis: Fundamentals and practices.
\newblock \emph{Ocean Modelling} 2018; 121: 49--75.

\bibitem{Chen2020}
Chen Y, Song L, Liu Y et~al.
\newblock A review of the artificial neural network models for water quality
  prediction.
\newblock \emph{Applied Sciences} 2020; 10(17): 5776.

\bibitem{Rousso2020}
Rousso BZ, Bertone E, Stewart R et~al.
\newblock A systematic literature review of forecasting and predictive models
  for cyanobacteria blooms in freshwater lakes.
\newblock \emph{Water Research} 2020; 182: 115959.

\bibitem{MIKE2022}
DHI.
\newblock {MIKE} {ECO} {L}ab, 2022.
\newblock
  \urlprefix\url{https://www.mikepoweredbydhi.com/products/mike-eco-lab}.
\newblock (Visited on 11/2022).

\bibitem{Delft2022}
Deltares.
\newblock {D}elft3{D} 4, 2022.
\newblock \urlprefix\url{https://oss.deltares.nl/web/delft3d}.
\newblock (Visited on 11/2022).

\bibitem{Velez2014}
V{\'e}lez C, Alfonso L, S{\'a}nchez A et~al.
\newblock Centinela: an early warning system for the water quality of the
  {C}auca river.
\newblock \emph{Journal of Hydroinformatics} 2014; 16(6): 1409--1424.

\bibitem{Silva2016}
Silva A, Pinto L, Rodrigues S et~al.
\newblock A {HAB} warning system for shellfish harvesting in {P}ortugal.
\newblock \emph{Harmful Algae} 2016; 53: 33--39.

\bibitem{Serramia2004}
Serrami{\'a} A.
\newblock {SAIH/SAICA}: automatic hydrological information system and automatic
  water quality information system in the {S}panish watersheds.
\newblock \emph{Tecno Ambiente (Espa{\~n}a)} 2004; 139: 15--18.

\bibitem{SERC2022}
{Shoutheast Environmental Research Center}.
\newblock {SERC} water quality monitoring network, 2022.
\newblock \urlprefix\url{http://serc.fiu.edu/wqmnetwork/}.
\newblock (Visited on 11/2022).

\bibitem{Riskin2021}
Riskin ML and Lee CJ.
\newblock {U.S.} {G}eological {S}urvey {N}ational {W}ater {Q}uality
  {M}onitoring {N}etwork.
\newblock Technical report, USGS National Water Quality Monitoring Network,
  Reston, VA, 2021.
\newblock \doi{10.3133/fs20213019}.

\bibitem{FWQMS2022}
ECCC.
\newblock Overview of freshwater quality monitoring and surveillance, 2022.
\newblock
  \urlprefix\url{https://www.canada.ca/en/environment-climate-change/services/freshwater-quality-monitoring/overview.html}.
\newblock (Visited on 11/2022).

\bibitem{Basha2008}
Basha EA, Ravela S and Rus D.
\newblock Model-based monitoring for early warning flood detection.
\newblock In \emph{Proceedings of the 6th ACM Conference on Embedded Network
  Sensor Systems}. SenSys '08, New York, NY, USA: Association for Computing
  Machinery.
\newblock ISBN 9781595939906, p. 295–308.
\newblock \doi{10.1145/1460412.1460442}.
\newblock \urlprefix\url{https://doi.org/10.1145/1460412.1460442}.

\bibitem{Curl2019}
Curl JM, Nading T, Hegger K et~al.
\newblock Digital twins: the next generation of water treatment technology.
\newblock \emph{Journal-American Water Works Association} 2019; 111(12):
  44--50.

\bibitem{Henares2022}
Henares K, Risco-Mart{\'\i}n JL, Ayala JL et~al.
\newblock Efficient micro data centres deployment for mobile healthcare
  monitoring systems in iot urban scenarios.
\newblock \emph{Journal of Simulation} 2022; : 1--15.

\bibitem{J-Cardenas2021}
Cárdenas R, Arroba P and Risco-Martín JL.
\newblock Bringing {AI} to the edge: A formal {M\&S} specification to deploy
  effective {IoT} architectures.
\newblock \emph{Journal of Simulation} 2021; 16(5): 494--511.
\newblock \doi{10.1080/17477778.2020.1863755}.

\bibitem{J-RiscoMartin2022b}
Risco-Martín JL, Mittal S, Henares K et~al.
\newblock x{DEVS}: {A} toolkit for interoperable modeling and simulation of
  formal discrete event systems.
\newblock \emph{Software: Practice and Experience} 2022; :
  1--42\doi{10.1002/spe.3168}.

\bibitem{J-RiscoMartin2022}
Risco-Martín JL, Henares K, Mittal S et~al.
\newblock A unified cloud-enabled discrete event parallel and distributed
  simulation architecture.
\newblock \emph{Simulation Modelling Practice and Theory} 2022; 118.

\bibitem{HerguedasPinedo2023}
Herguedas-Pinedo B, Risco-Martín JL, Esteban S et~al.
\newblock Predictive modeling and simulation system for the management of
  harmful cyanobacteria blooms.
\newblock In \emph{Proceedings of the 2023 Annual Modeling and Simulation
  Conference (ANNSIM'23)}.

\bibitem{Video2022}
Esteban S.
\newblock {HAB} {M}onitoring {S}imulation {V}ideo, 2022.
\newblock \urlprefix\url{https://archive.org/details/habsim-20220920}.

\bibitem{Niyonkuru2021}
Niyonkuru D and Wainer G.
\newblock A {DEVS}-based engine for building digital quadruplets.
\newblock \emph{SIMULATION} 2021; 97(7): 485--506.
\newblock \doi{10.1177/00375497211003130}.
\newblock \urlprefix\url{https://doi.org/10.1177/00375497211003130}.
\newblock PMID: 34219819, \eprint{https://doi.org/10.1177/00375497211003130}.

\bibitem{FerreroLosada2023}
Ferrero-Losada S, Besada-Portas E, Risco-Martín JL et~al.
\newblock D{EVS}-based modeling and simulation of data-driven exploration
  algorithms of lentic water bodies with an {ASV}.
\newblock In \emph{Proceedings of the 2023 Annual Modeling and Simulation
  Conference (ANNSIM'23)}.

\end{thebibliography}
